\documentclass[showpacs,aps,twocolumn,floatfix,prb,longbibliography, superscriptaddress]{revtex4-2}
\usepackage{bm,amsmath,amssymb, mathrsfs}
\usepackage{bbm} %for unit matrix \mathbbm{1}
\usepackage{physics}
\usepackage{enumerate}
\usepackage{commath}
\usepackage{braket} %for Bra-Ket notation
\usepackage{graphicx,bm}
\usepackage{verbatim}
\usepackage[dvipsnames]{xcolor}
\usepackage{soul}
\usepackage[separate-uncertainty = true]{siunitx}
\usepackage[colorlinks]{hyperref}
\AtBeginDocument{%
    \hypersetup{    
        linkcolor=blue,
        filecolor=magenta,      
        urlcolor=blue,
        citecolor=blue,
     }
}

\newcommand{\eu}{\mathrm{e}}
\newcommand{\imi}{\mathrm{i}}

\usepackage[T1]{fontenc}
\usepackage[utf8]{inputenc}

\begin{document}

\title{Cooperativity-enhanced single-photon emission in a topological quantum emitter chain}

\author{Jonathan Sturm}
\email{jonathan.sturm@uni-wuerzburg.de}
\affiliation{Julius-Maximilians-Universit{\"a}t W{\"u}rzburg, Institute for Theoretical Physics and Astrophysics and W{\"u}rzburg-Dresden Cluster of Excellence ctd.qmat, Am Hubland, 97074 W{\"u}rzburg, Germany}

\author{Irene Burghardt}
\affiliation{Goethe University Frankfurt, Institute of Physical and Theoretical Chemistry, Max-von-Laue-Str.\ 7, 60438 Frankfurt, Germany}

\author{Bert Hecht}
\affiliation{Julius-Maximilians-Universität Würzburg, Nano-Optics \& Biophotonics Group, Experimental Physics 5, Am Hubland, 97074 W{\"u}rzburg, Germany}

\author{Adriana P\'alffy}
\email{adriana.palffy-buss@uni-wuerzburg.de}
\affiliation{Julius-Maximilians-Universit{\"a}t W{\"u}rzburg, Institute for Theoretical Physics and Astrophysics and W{\"u}rzburg-Dresden Cluster of Excellence ctd.qmat, Am Hubland, 97074 W{\"u}rzburg, Germany}

\date{\today}% It is always \today, today,

\begin{abstract}
Single-photon emitters are one of the key elements in state-of-the art quantum technologies, yet optimizing and manipulating their emission requires precise fabrication of the emitter systems themselves. 
Here, we propose a route to tailored single-photon emission by exploiting cooperative effects in a topological single-photon emitter chain. 
Although the involvement of multiple emitters normally suppresses single-photon emission, we show that localized topological edge modes restore single-photon emission and protect against local perturbations. 
At the same time, cooperative effects occur due to the collective nature of the edge modes, leading to directional emission and linewidth narrowing. 
Our findings open new possibilities for tailoring single-photon emission by combining cooperativity and topological effects.
\end{abstract}

\maketitle

\section{Introduction}
Single-photon emitters (SPEs) play a central role in state-of-the-art quantum technologies such as quantum computing \cite{Obrien2007}, cryptography \cite{Scarani2009, Pirandola2020}, or quantum communication \cite{Vajner2022, Azuma2023, Lounis2005, Keni2025}.
By now, SPEs have been realized in a plethora of physical platforms, from individual atoms \cite{Kimble1977} and color centers \cite{Kurtsiefer2000, Aharonovich2011} over mesoscale systems like molecules \cite{Brunel1999} and quantum dots \cite{Lodahl2015, Ding2016, Kagan2021, Hornung2024} to many-body systems like atomic vapors and Rydberg super-atoms \cite{Keni2025}.
Near-perfect SPEs providing on-demand indistinguishable photons one at a time have been demonstrated in semiconductor quantum dots \cite{Ding2016, Hornung2024}.

SPEs are a special class of quantum emitters, which are often modeled as two-level systems that can be excited with light and emit by spontaneous decay of the excited state \cite{Hecht_book}. 
Multiple quantum emitters interacting with a common photonic environment form collective states maintained by dipole-dipole couplings and collective decay \cite{Lehmberg1970I, Lehmberg1970II, Jen2025}. 
Such collective states can lead to cooperative emission phenomena like super- and subradiance \cite{Dicke1954, Garcia2017, Reitz2022}, directional emission \cite{Bychek2025, Tonoyan2026}, and lasing \cite{Holzinger2020}. 
However, due to the simultaneous involvement of multiple emitters, cooperativity usually prohibits single-photon emission.

In this work, we show that a topological quantum emitter chain enables single-photon emission in a cooperative state. 
To this end, we consider a chain of quantum emitters in a Su-Schrieffer-Heeger geometry - a dimerized model that provides topologically protected localized edge eigenstates \cite{Su1979, Asboth_book} - and incoherently pump one edge emitter.
Even for this generic pumping scheme, which does not consider specifications like laser detuning or polarization, we show that  topological edge modes can be excited in the steady state. 
While the strong localization of the edge modes restricts emission to a single edge emitter and thereby restores single-photon emission, the superpositional nature of edge modes opens up multiple interference paths enabling cooperative effects like directional emission and linewidth narrowing. 
Since the number of topological edge states depends on whether the chain has an even or odd number of sites, adding or removing one edge emitter drastically alters the radiation pattern.
Due to topological protection, the observed single-photon emission is robust against disorder, which is likely to occur in experimental realizations, as long as the edge states are conserved.
Our work paves the way for topologically protected SPEs whose properties can be altered or controlled by cooperative effects.

\begin{figure}
    \centering
    \includegraphics[width=\linewidth]{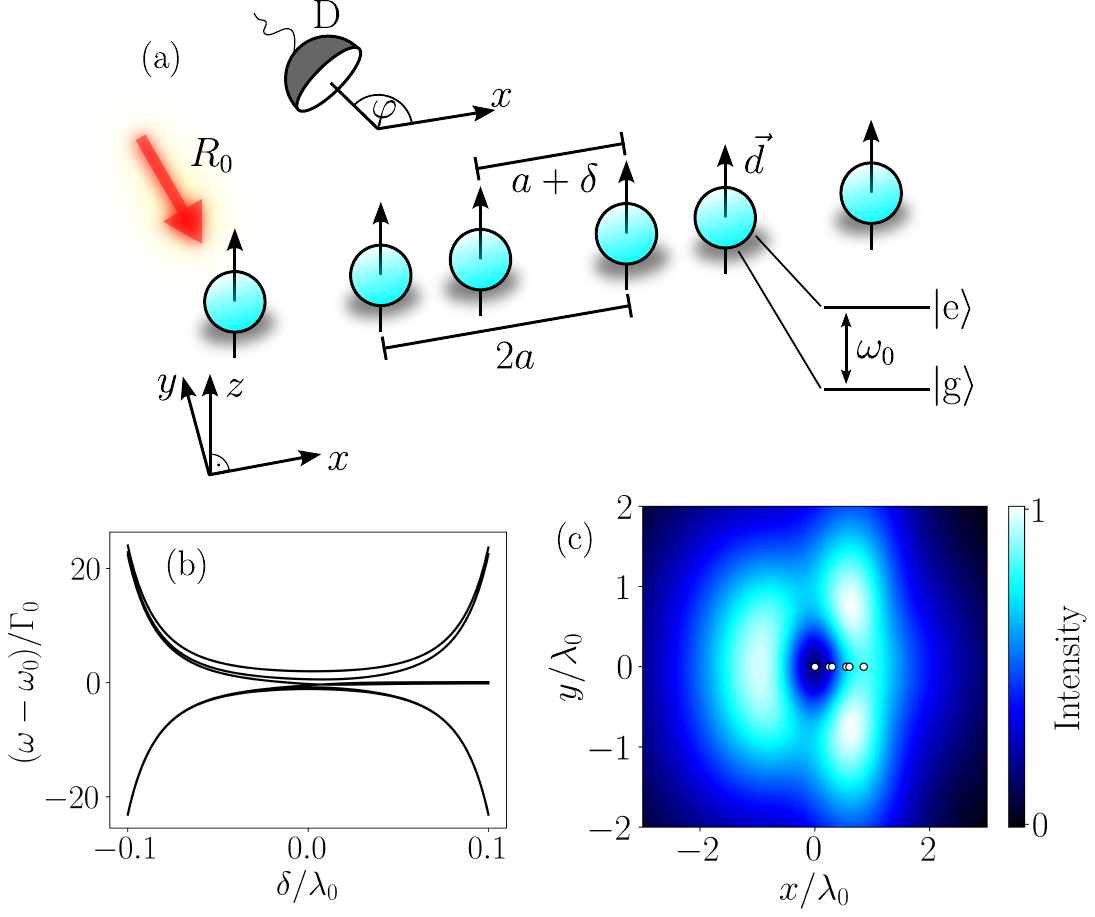}
    \caption{Quantum emitter SSH chain.
    (a) Sketch of the setup under consideration. 
    The cyan dots are quantum emitters with ground and excited states $\ket{\mathrm{g/e}}$ separated by a transition frequency $\omega_0$ with dipole moment $\vec{d} \parallel \hat{z}$.
    The emitters form a dimerized chain along the $x$ axis ($y,z=0$) with lattice constant $2a$ and dimerization parameter $\delta$.
    The first emitter from the left is incoherently pumped with pumping rate $R_0$ (red arrow). A detector D is positioned within the $x$-$y$ plane at an angle $\varphi$ with respect to the $x$ axis.
    (b) Eigenvalues of the Hamiltonian in Eq.~\eqref{eq Hamiltonian} for varying $\delta$. 
    (c) Calculated intensity in arbitrary units in the plane $z=\lambda_0$ ($\lambda_0 = 2\pi/k_0$) for $\delta=0.1\lambda$ and $R_0 = \Gamma_0$.
    White dots are the emitter positions.
    In (b) and (c) we used $N=6$ and $a=0.15\lambda_0$. 
    }
    \label{fig figure 1}
\end{figure}
\section{Model}
We consider a one-dimensional dimerized chain of $N$ quantum emitters shown in Fig.~\ref{fig figure 1}(a) with ground and excited states $\ket{\mathrm{g}_i}$ and $\ket{\mathrm{e}_i}$ ($i=1,...,N$), respectively. 
The emitters are distributed along the $x$ axis with positions $\vec{r}_i = x_i \hat{x}$ with
\begin{equation}
    x_i = 
    \begin{cases}
        (i-1)a & (i \ \mathrm{odd}), \\
        (i-1)a + \delta & (i \ \mathrm{even}),
    \end{cases}
\end{equation}
for $i=1,...,N$. 
Here, $a$ determines the inter-emitter spacing and $-a<\delta<a$ induces a dimerization of the chain. 
Significant cooperative effects require $a<\lambda_0$ where $\lambda_0$ is the transition wavelength \cite{Garcia2017}.

Following the theory of open quantum systems, the time-evolution of the density matrix $\varrho$ of the chain under the Born and Markov approximations is governed by the master equation \cite{Breuer_book, Garcia2017}
\begin{equation}\label{eq master equation}
    \dot{\varrho} 
    = -\imi [H, \varrho] + \mathcal{D}[\varrho].
\end{equation}
The coherent part of the time evolution is described by the Hamiltonian \cite{Garcia2017, Reitz2022}
\begin{equation}\label{eq Hamiltonian}
    H = \sum_{i=1}^N \omega_0 \sigma_i^\dagger\sigma_i +  \sum_{\substack{i,j=1 \\ i\neq j}}^N \Omega_{ij} \sigma_i^\dagger\sigma_j
\end{equation}
with transition operators $\sigma_i = \ketbra{\mathrm{g}_i}{\mathrm{e}_i}$ and transition frequency $\omega_0$ as well as dipole-dipole couplings $\Omega_{ij}$ between excited states $i$ and $j$.
The dissipator consists of the Lindblad operators $\mathcal{D} = \mathscr{L}_0 + \mathscr{L}_{\rm pump}$, where $\mathscr{L}_{\rm pump}$ models incoherent pumping (see below) and 
\begin{equation}\label{eq lindblad operator}
    \mathscr{L}_0[\varrho] = \sum_{i,j=1}^N \frac{\Gamma_{ij}}{2} \qty(2\sigma_i\varrho\sigma_j^\dagger - \sigma_i^\dagger\sigma_j\varrho - \varrho\sigma_i^\dagger\sigma_j)
\end{equation}
is the (non-local) collective decay with rates $\Gamma_{ij}$ \cite{Garcia2017, Reitz2022}.
The dipole-dipole couplings $\Omega_{ij}$ and the decay rates $\Gamma_{ij}$ are related to the transition dipole moments $\vec{d}_i$ of each atom by \cite{Garcia2017, Dung2002}
\begin{equation} 
    \Omega_{ij} - \imi\frac{\Gamma_{ij}}{2}
    = -\Gamma_0 \frac{3\pi}{k_0} \vec{d}_i \cdot G(\vec{r}_i-\vec{r}_j, \omega_0) \cdot \vec{d}_j,
\end{equation}
where $\Gamma_0 = \Gamma_{ii}$ is the local decay rate and $k_0 = \omega_0/c$ is the wave number, with $c$ being the speed of light. 
The Green's function \cite{Hecht_book, Garcia2017}
\begin{align}\label{eq greens function}
    G(\vec{r}, \omega_0) =  \frac{\eu^{\imi k_0 r}}{4\pi k_0^2 r^3} \Big( 
    & (k_0^2r^2 + \imi k_0r - 1)\mathbbm{1} \nonumber \\
    & - (k_0^2 r^2 + 3\imi k_0r - 3) \hat{r}\otimes\hat{r} \Big)
\end{align}
describes the (classical) propagation of light between a dipole source at $\vec{r}_j$ to a dipole receiver at $\vec{r}_i$.

The dimerization induces alternating couplings such that the Hamiltonian in Eq.~\eqref{eq Hamiltonian} can be viewed as a quantum optical realization of the Su-Schrieffer-Heeger (SSH) Hamiltonian \cite{Wang2018, Sturm2025}, which is known for having two distinct topological phases, one of which entails the existence of localized edge eigenstates at zero energy. 
By diagonalizing the matrix elements of the Hamiltonian, $\omega_0\delta_{ij} + \Omega_{ij}$, for varying dimerization $\delta$, we obtain the band structure shown in Fig.~\ref{fig figure 1}(b).
In the topological phase ($\delta>0$), we observe two eigenstates with eigenfrequency $\omega=\omega_0$, which are absent for $\delta<0$.

Since we are interested in studying emission from the topological edge state, we consider incoherent pumping of the left-edge emitter only:
\begin{equation}
    \mathscr{L}_{\rm pump}[\varrho] = \frac{R_0}{2} \qty(2\sigma_1^\dagger\varrho\sigma_1 - \sigma_1\sigma_1^\dagger\varrho - \varrho\sigma_1\sigma_1^\dagger),
\end{equation}
where $R_0$ is the pump rate.
Starting from the collective ground state, we let the density matrix $\varrho$ evolve according to Eq.~\eqref{eq master equation} until a steady state $\varrho_{\rm ss}$ is reached \cite{quantumpotics}.
In the steady state, we analyze the emitted light in terms of the intensity distribution \cite{Glauber_book}
\begin{equation}
    I(\vec{r}) = \braket{\vec{E}^\dagger(\vec{r},0) \cdot\vec{E}(\vec{r},0)}_{\rm ss},
\end{equation}
and the power spectrum \cite{Glauber_book, Carmichael_book}
\begin{equation}
    S(\omega, \vec{r}) = 2\mathrm{Re} \int_0^\infty \braket{\vec{E}^\dagger(\vec{r}, \tau) \cdot \vec{E}(\vec{r}, 0)}_{\rm ss} \eu^{-\imi\omega\tau} \dd{\tau}.
\end{equation}
Here, $\braket{A}_{\rm ss} = \tr(\varrho_{\rm ss} A)$ is the expectation value of an operator $A$ and 
\begin{equation}
    \vec{E}(\vec{r}, \tau) = \mu_0\omega_0^2 \sum_{i=1}^N G(\vec{r}-\vec{r}_i, \omega_0) \cdot \vec{d}_i \, \sigma_i(\tau)
\end{equation}
is the field annihilation operator at time $\tau$ and position $\vec{r}$, and $\mu_0$ is the vacuum permeability \cite{Garcia2017}.
As an example, the intensity distribution for a chain of $N=6$ atoms is shown in Fig.~\ref{fig figure 1}(c).

In the following, we will always assume identical dipole moments $\vec{d}_i = |\vec{d}| \hat{z}$ perpendicular to the chain axis for all emitters $i=1,...,N$. 
Since this arrangement mainly emits into the the $x$-$y$ plane, we focus on emission into this plane and parametrize $\vec{r} = r(\cos(\varphi)\hat{x} + \sin(\varphi)\hat{y})$, see Fig.~\ref{fig figure 1}(a).
In the far-field limit $r/\lambda_0\gg1$, observables become independent of $r$ such that $I(\vec{r}) \equiv I(\varphi)$ and $S(\omega,\vec{r}) \equiv S(\omega, \varphi)$.
In this limit, we also study the second-order correlation function \cite{Glauber_book}
\begin{equation}
    g^{(2)}(0) = \frac{\braket{\vec{E}^\dagger(\vec{r}) \cdot (\vec{E}^\dagger(\vec{r})\cdot\vec{E}(\vec{r})) \vec{E}(\vec{r})}_{\rm ss}}{(\braket{\vec{E}^\dagger(\vec{r}) \cdot \vec{E}(\vec{r})}_{\rm ss})^2}
\end{equation}
to distinguish classical from quantum light.
Explicit expressions for the far-field limits are derived in Appendix~\ref{appendix far-field}.

\section{Single-photon emission}
\begin{figure}
    \centering
    \includegraphics[width=\linewidth]{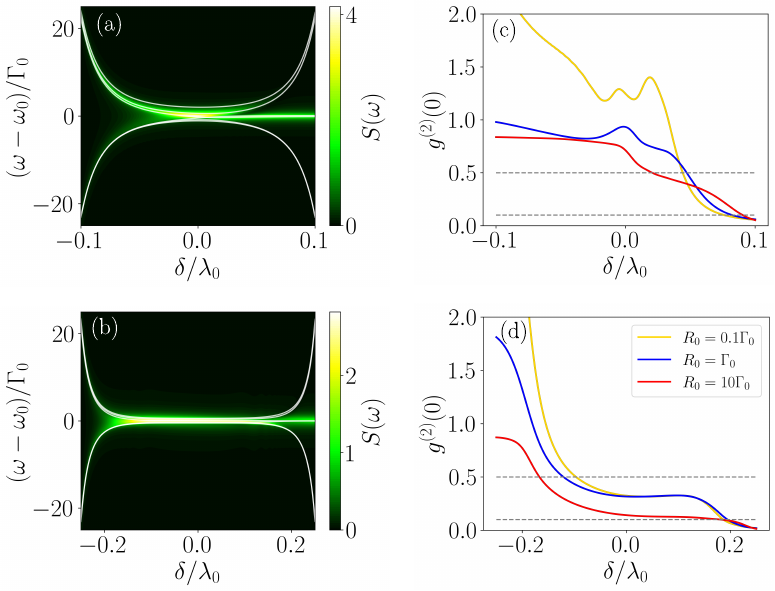}
    \caption{(a) Power spectrum $S(\omega)$ for varying $\delta$ with $a=0.15\lambda_0$ and $R_0 = \Gamma_0$ computed in the far-field limit for $\varphi=\pi$.
    White lines are the single-excitation eigenenergies of the dipole-dipole Hamiltonian, cf. Fig.~\ref{fig figure 1}(b).
    (b) Same as (a) for $a=0.3\lambda_0$.
    (c) Second-order correlation function for $a=0.15\lambda_0$ and varying  pump rate $R_0$. 
    The upper gray dashed line is the single-photon border $g^{(2)}(0) = 0.5$, the lower one marks the limit of $g^{(2)}(0) = 0.1$.
    (d) Same as (c) for $a=0.3\lambda_0$. 
    The chain length in all panels is $N=6$.
    }
    \label{fig figure 2}
\end{figure}
To determine whether our pump scheme is capable of exciting the edge modes in the topological phase, we compute the power spectrum $S(\omega)$ for varying $\delta$.
The results are shown in Figs.~\ref{fig figure 2}(a) and (b) for $a=0.15\lambda_0$ and $a=0.3\lambda_0$, respectively.
The white lines are the single-excitation eigenenergies of the dipole-dipole Hamiltonian in Eq.~\eqref{eq Hamiltonian}, cf. Fig.~\ref{fig figure 1}(b), and serve as orientation. 
States with non-zero power spectrum contribute to the steady state and emission. 
In the topological phase ($\delta>0$), emission occurs almost entirely via the zero-frequency states, proving that by pumping the edge, the edge states are populated.
This is confirmed in Appendix~\ref{appendix edge states}, where the steady-state populations are calculated in the eigenbasis.

Our next step is to calculate the second-order correlation function $g^{(2)}(0)$ for varying $\delta$ and $R_0$ to determine the quantum nature of the emitted light. 
The results are shown in Figs.~\ref{fig figure 2}(c) and (d) for $a=0.15\lambda_0$ and $a=0.3\lambda_0$, respectively. 
In Fig.~\ref{fig figure 2}(c), we see that for $R_0=\Gamma_0/10$ (yellow curve), in the non-topological phase ($\delta<0$) we have $g^{(2)}(0) >1$, indicating thermal (classical) light. 
At around $\delta\approx 0.05\lambda_0$, the curve falls below 1 (coherent light). 
For $\delta>0.05\lambda_0$, the single-photon regime $g^{(2)}(0) <0.5$ is reached \cite{Keni2025}.
Perfect single-photon emission $g^{(2)}(0) = 0$ is obtained in the deep topological phase for $\delta=0.1\lambda_0$.
For the higher pump rates $R_0=\Gamma_0$ and $R_0=10\Gamma_0$ (blue and red curves, respectively) we observe mainly the same behavior, except already for $\delta<0$ the correlation is below 1.

Similar patterns can be observed for the case $a=0.3\lambda_0$ shown in Fig.~\ref{fig figure 2}(d), although the correlation is generally lower than for $a=0.15\lambda_0$.
This is due to the overall smaller couplings between the atoms, which depend roughly on $1/r^3$ if $r$ is the distance between two atoms, cf. Eq.~\eqref{eq greens function}.
Besides that, all three curves fall below the $g^{(2)}(0) = 0.5$ mark already in the non-topological phase.
The single-photon limit $g^{(2)}(0)=0$ is again reached in the deep topological phase at $\delta\geq0.2\lambda_0$.
In conclusion, we have shown both that we can excite the edge states by incoherently pumping the edge and that the edge state produces single-photon emission.

\section{Far-field Intensity}
\begin{figure}
    \centering
    \includegraphics[width=1\linewidth]{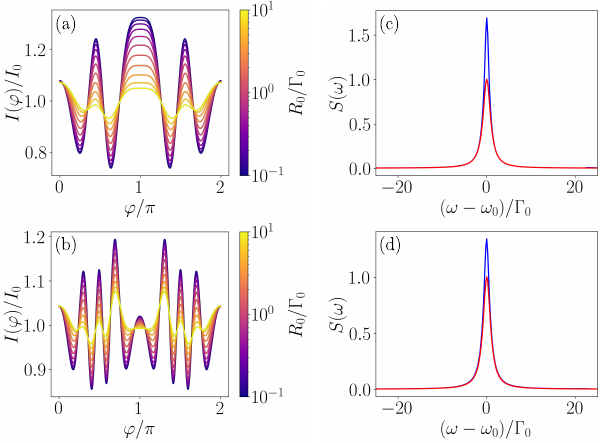}
    \caption{(a) Far-field intensity $I$ as function of angle $\varphi$ normalized by the (isotropic) single-emitter intensity $I_0$ for $a=0.15\lambda_0$ and $\delta = 0.1\lambda_0$. 
    (b) Same as (a) for $a=0.3\lambda_0$ and $\delta=0.25\lambda_0$
    (c) Far-field power spectrum $S(\omega)$ in direction $\varphi=\pi$ for $a=0.15\lambda_0$, $\delta=0.1\lambda_0$, and $R_0=\Gamma_0$ (blue curve).
    Red curve shows single-emitter spectrum.
    (d) Same as (c) for $a=0.3\lambda_0$, $\delta=0.25\lambda_0$, and $\varphi=0.7\pi$.
    In all panels we used $N=6$.
    }
    \label{fig figure 3}
\end{figure}
Having shown that the topological edge state restores single-photon emission, we are now going to investigate how cooperativity changes emission properties compared to a single quantum emitter.
For that purpose, we first compute the far-field intensity as a function of angle $\varphi$, the results of which are shown in Figs.~\ref{fig figure 3}(a) and (b) for $(a, \, \delta) = (0.15\lambda_0, \, 0.1\lambda_0)$ and $(a, \, \delta) = (0.3\lambda_0, \, 0.25\lambda_0)$, respectively.
While the single-emitter intensity $I_0$ is always isotropic, emission from the chain strongly depends on the emission angle. 
This $\varphi$ dependence is due to the interference between individual quantum emitters, which is incorporated in the coherences $\braket{\sigma_i^\dagger\sigma_j}_{\rm ss}$ in the intensity,
\begin{equation}\label{eq Intensity phi}
    I(\varphi) \sim \sum_{i,j=1}^N \eu^{\imi k_0 (x_i-x_j)\cos(\varphi)} \braket{\sigma_i^\dagger \sigma_j}_{\rm ss},
\end{equation}
see also Appendix~\ref{appendix far-field}.
Due to this interference, we observe intensity gains up to $\approx30\%$ ($a=0.15\lambda_0$) and $\approx20\%$ ($a=0.3\lambda_0$), depending on the pump rate.
This effect is similar to the formation of lobes in the emission pattern of classical coupled-antenna systems such as Yagi-Uda antennas \cite{Hecht_book}.

Intensity gains can also be seen in the power spectrum, which is computed in Figs.~\ref{fig figure 3}(c) and (d) for $a=0.15\lambda_0$ and $a=0.3\lambda_0$, respectively, in the direction of maximum emission. 
The peak of the blue curve (spectrum of the chain) exceeds the one of the red curve (single emitter) by $\approx68\%$ ($a=0.15\lambda_0$) and $\approx34\%$ ($a=0.3\lambda_0$), narrowing the linewidth (full width at half maximum) by $\approx20\%$ and $\approx10\%$, respectively.
We note that the observed linewidth narrowing is not to be confused with the one reported in Ref.~\cite{Wang2024}, which studied the eigenvalue broadening due to local noise in the SSH model, while the linewidth considered here originates from spontaneous decay.

\section{Single-excitation subspace}
\begin{figure}
    \centering
    \includegraphics[width=\linewidth]{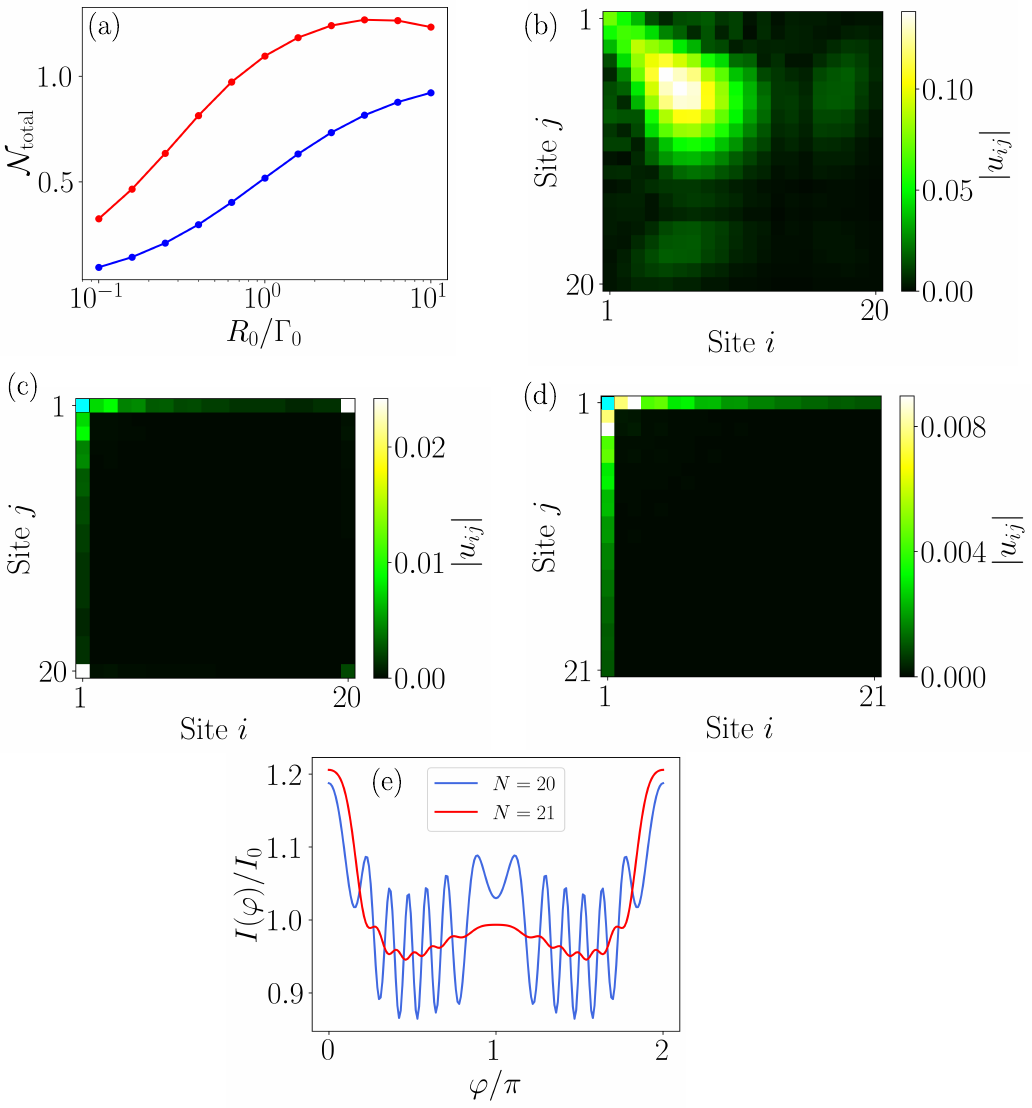}
    \caption{(a) Total population of a chain with $N=6$ emitters for $\delta=0.1\lambda_0$ (blue curve) and $\delta=0$ (red curve).
    (b) Absolute values of the coherences and populations $|u_{ij}| = |\braket{\sigma_i^\dagger\sigma_j}_{\rm ss}|$ for $N=20$ and $\delta=0$.
    (c) Same as (b) for $\delta=0.1\lambda_0$.
    The first entry (cyan) has the value $|u_{11}|\approx 0.5$.
    (d) Same as (c) for $N=21$.
    (e) Far-field intensity $I(\varphi)$ normalized by the single-emitter intensity $I_0$ for $\delta=0.1\lambda_0$.
    Parameters are $a=0.15\lambda_0$ and in (b)-(e) $R_0=\Gamma_0$.
    Panels (b)-(e) are computed in the single-excitation subspace.
    }
    \label{fig figure 4}
\end{figure}
Since the interference pattern in the intensity depends on the coherences and populations $u_{ij} = \braket{\sigma_i^\dagger\sigma_j}_{\rm ss}$, see Eq.~\eqref{eq Intensity phi}, we are now going to further analyze how these quantities are related to the topological edge states. 
The total population of the chain is given by
\begin{equation}
    \mathcal{N}_{\rm total} = \sum_{i=1}^N u_{ii},
\end{equation}
which we show for both a trivial chain with $\delta=0$ (red curve) and a topological chain (blue curve) in Fig.~\ref{fig figure 4}(a).
One can see that the blue curve never exceeds a total excitation of 1, while the trivial chain only slightly exceeds this limit for pump rates $R_0/\Gamma_0 \geq 1$.
This allows us to switch to the single-excitation subspace defined by the basis $\qty{\ket{\mathrm{GS}}, \, (\ket{i})_{i=1,...,N}}$ where $\ket{\mathrm{GS}}$ is the collective ground state and $\ket{i} = \sigma_i^\dagger\ket{\mathrm{GS}}$ is the state where emitter $i$ is excited and all other emitters are in the ground state.
Within this regime, we are able to study chain lengths that are untractable using the full Hilbert space.

We begin our analysis of the coherences $u_{ij}$ by considering a topologically trivial chain ($\delta=0$) as the results here are rather intuitive and serve as a valuable contrast to the topological case. 
The absolute values $|u_{ij}|$ for a chain of $N=20$ emitters are shown in Fig.~\ref{fig figure 4}(b).
Recalling that the diagonal elements $|u_{ii}|$ are the populations of the individual emitters, we can see that from the first emitter the excitation propagates into the bulk of the chain up to a certain depth (approximately half of the chain length). 
At this point, due to spontaneous emission, on average no excitation is transmitted further into the bulk.

In the topological case shown in Fig.~\ref{fig figure 4}(c), the situation looks very different. 
While the population of the first emitter (cyan) is $|u_{11}|\approx 0.5$, the rest of the chain is in the ground state. 
However, although excitation does not enter the bulk of the chain, the coherences $|u_{1i}| = |u_{i1}|$, which link the first emitter to the rest of the chain, have small but finite values. 
Strikingly, the highest value is obtained for $|u_{1N}| = |u_{N1}|$, which connects the first and the last atoms. 
This is a consequence of the topological edge modes $\ket{\psi_\pm}$, which  are roughly symmetric and antisymmetric superpositions of the edge emitters:
\begin{equation}\label{eq edge states}
    \ket{\psi_\pm} \approx \frac{1}{\sqrt{2}} (\ket{N} \pm \ket{1}),
\end{equation}
see also Appendix~\ref{appendix edge states}.
A close look also reveals that the last emitter actually has a finite population of $|u_{NN}| \approx 0.23\%$, which is significantly larger than the total population of the remaining bulk part of the chain, $\sum_{i=2}^{N-1} u_{ii} \approx 0.09\%$.
This anomalous population transfer from the first to the last atom without populating the bulk clearly is a consequence of the non-local wavefunction of the topological edge modes in Eq.~\eqref{eq edge states}.

A consequence of the comparatively high values of the coherences $|u_{1N}|$ and $|u_{N1}|$ can be seen in the far-field intensity. 
From Eq.~\eqref{eq Intensity phi} we know that the coherences serve as a weight for a phase factor which contains the emission angle $\varphi$ and the distance between the atoms $x_i - x_j$, encoding interference between individual emission channels.
The large distance between the edge atoms leads to fast oscillation in the phase factor, which becomes visible for a sufficiently large weighting factor $|u_{1N}|$.
These oscillations can be seen in the blue intensity plot in Fig.~\ref{fig figure 4}(e).
To prove that these oscillations originate from the superposition of the edge emitters, we compare our results to a chain with an odd number ($N=21$) emitters. 
In this case, there is only one edge state $\ket{\psi} \approx \ket{1}$ such that there are no anomalous values for the coherences $|u_{1N}|$ and $|u_{N1}|$ anymore and the last emitter has zero population, see Fig.~\ref{fig figure 4}(d).
For that reason, the corresponding intensity in the red curve in Fig.~\ref{fig figure 4}(e) shows no significant oscillations.
Similar patterns distinguishing even and odd chain lengths have been discussed in the context of higher-harmonic generation in Ref.~\cite{Bauer2018}.

\section{Disorder}
\begin{figure}
    \centering
    \includegraphics[width=\linewidth]{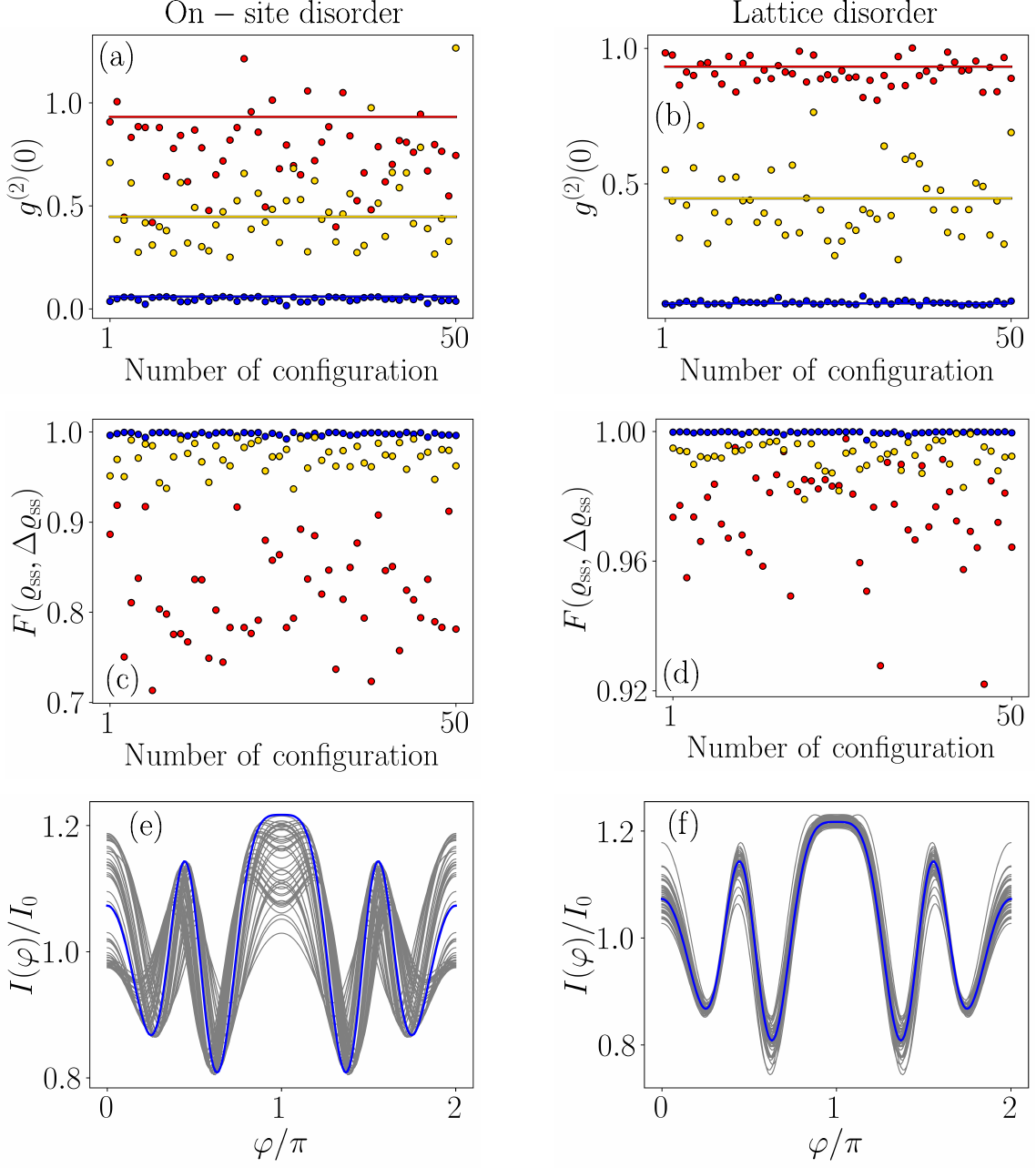}
    \caption{Influence of on-site (left column) and lattice disorder (right column) for a chain with $N=6$ emitters and $a=0.15\lambda_0$ with $R_0=\Gamma_0$. 
    (a) and (b) show $g^{(2)}(0)$ for $\varphi=\pi$ and $\delta=0$ (red), $\delta=0.05\lambda_0$ (yellow), and $\delta=0.1\lambda_0$ (blue dots). 
    The lines show the values without disorder. 
    (c) and (d) show the fidelity $F$ (same color code as above).
    (e) and (f) show the intensity as function of $\varphi$ with (gray lines) and without disorder (blue curves) normalized by the single-emitter value $I_0$.
    We have calculated $50$ configurations of random disorder with $\Delta\omega_i$ and $\Delta x_i$ having Gaussian distributions with standard deviations of $1\Gamma_0$ and $0.005\lambda_0$, respectively.}
    \label{fig figure 5}
\end{figure}
One of the most prominent features of topological edge modes is that they are protected against local disorder if (1) the disorder strength does not exceed the topological gap and (2) disorder does not break the fundamental symmetries of the Hamiltonian.
We consider two types of local disorder: variations in the transition frequency (``on-site disorder'') modeled by a perturbative Hamiltonian 
\begin{equation}
    \Delta H = \sum_{i=1}^N \Delta\omega_i \, \sigma_i^\dagger\sigma_i,
\end{equation}
such that the Hamiltonian in Eq.~\eqref{eq Hamiltonian} becomes $H \to H+\Delta H$, and variations in the emitter positions (``lattice disorder''), 
\begin{equation}
    x_i \to x_i + \Delta x_i.
\end{equation}
Note that on-site disorder breaks the chiral symmetry of the SSH model, leading to asymmetries in the eigenvalue spectrum and a shift of the edge-state frequencies away from $\omega_0$. 
However, as long as the edge states stay well-separated from the bulk states, the localization is preserved.

We show the results of our disorder analysis in Fig.~\ref{fig figure 5}.
In Figs.~\ref{fig figure 5}(a) and (b) we compute the $g^{(2)}(0)$ correlation function for 50 random on-site and lattice disorder configurations, respectively. 
We distinguish between three cases, the trivial phase ($\delta=0$, red), the shallow topological phase ($\delta=0.05\lambda_0$, yellow), and the deep topological phase ($\delta=0.1\lambda_0$, red dots), which we compare to their values without disorder (continuous lines). 
In both the trivial and the shallow topological phase, the correlation is broadly scattered around the ideal values since for $\delta=0.05\lambda_0$ the disorder strength is of the same order as the topological gap (for $\delta=0$ there is no gap at all).
In contrast to this, for $\delta=0.1\lambda_0$ the large topological gap protects the edge modes against both on-site and lattice disorder, such that we observe only marginal variations of $g^{(2)}(0)$ around the ideal value in the single-photon regime.

A similar picture is obtained when directly comparing the disordered steady state $\Delta\varrho_{\rm ss}$ with the steady state without disorder $\varrho_{\rm ss}$ by computing the quantum fidelity \cite{Campaioli2024}
\begin{equation}
    F(\varrho_{\rm ss}, \, \Delta\varrho_{\rm ss})
    = \tr\qty(\sqrt{\sqrt{\varrho_{\rm ss}} \Delta\varrho_{\rm ss} \sqrt{\varrho_{\rm ss}}})^2.
\end{equation}
In Figs.~\ref{fig figure 5}(c) and (d) we again observe that the deep topological phase remains stable against both types of perturbation, while in the trivial and shallow topological phases the steady state for many configurations shows significant losses in fidelity.

In Figs.~\ref{fig figure 5}(e) and (f), we compare the directional intensity $I(\varphi)$ of the topological chain ($\delta=0.1\lambda_0$) with (gray lines) and without disorder (blue curve). 
Figure~\ref{fig figure 5}(e) shows that on-site disorder has quite significant influence on the intensity, although the quantum state fidelity is close to 1. 
As seen in our discussion of Fig.~\ref{fig figure 4}(c), the intensity distribution mainly depends on four entries of the coherence matrix, $u_{11}$, $u_{NN}$, and $u_{1N} = u_{N1}^*$.
Therefore, although changes in these four values have little impact on the fidelity (which takes into account the whole $2^N \times 2^N$-sized density matrix), the intensity can yet show visible changes. 
In contrast to this, these four values are quite robust against deformations of the lattice, which leave the topological edge modes untouched.
Therefore, we only observe small changes in the intensity in Fig.~\ref{fig figure 5}(f).

\section{Experimental platforms}
As stated in the introduction, there is a large number of possible realizations of quantum emitters, yet for our system one has to be able to position the emitters in a regular lattice with spacings down to $0.05\lambda_0$.
For optical wavelengths, this corresponds to spacings between \SI{10}{nm} and \SI{100}{nm}.

Experimental realization of chains of emitters with spacings on the order of \SI{100}{nm} is a difficult but generally possible task. 
For example, quantum dots can be positioned using holes written into photoresist and subsequent lift-off at very high precision down to the \SI{10}{nm} scale \cite{Xie2015, kong2025}.
Short subwavelength chains of quantum dots have been successfully implemented to achieve unidirectional emission \cite{Curto2010}.
However, since this method entails an element of randomness as to whether a site is filled or not, reliable fabrication of longer chains becomes significantly more challenging.

Alternatively, one may consider 1D molecular J-aggregates \cite{Wuerthner2011,Hestand2018} which can be carefully cut into separate pieces using, e.g., He-ion beam milling or electron beam irradiation \cite{Seo2018}. 
Each separate piece of J-aggregate then effectively acts as a single emitter while the coupling between these emitters is modulated by the width of the cut gap. 
An important aspect in these molecular structures is the role of decoherence beyond spontaneous decay described by the Lindblad operator in Eq.~\eqref{eq lindblad operator}. Such additional processes induced by the environment or phonons could be significant due to the large spatial separation between the edge emitters and the low excitation of the bulk emitters lying in between \cite{Schlosshauer_book}.

Further methods include the use of hybrid quantum systems that consist of quantum dots coupled to plasmonic resonators. 
Topological SSH chains made from plasmonic resonator chains \cite{Schurr2025} could be used as scaffolds in which photopolimerization of material in local field hotspots could be used to deterministically deposition quantum dots in respective sites. However, similar arguments as above apply with respect to the efficiency to produce larger systems. 
Plasmonic SHH chains without quantum emitters may serve as classical counterparts in which coherence effects predicted in this study could be tested to some degree, e.g., regarding their antenna properties.

\section{Conclusion}
We have shown that edge states of a topological quantum emitter SSH chain can be excited by incoherently pumping the edge emitter. 
Due to the strong localization of the edge state, we observe single-photon emission in the steady state, while at the same time cooperativity changes the emission spectra and intensities. 
Although the emission pattern shows sensitivity against local perturbations, single-photon emission is protected by the topological band gap. 
Moreover, we have shown that the superpositional nature of the edge modes in chains with an even number of emitters leads to fast oscillations in the radiation pattern. 
Our results prove that topology offers a way to observe cooperative effects while maintaining single-photon emission, enabling tailored emission without altering the individual emitter.

\begin{acknowledgments}
The authors gratefully
acknowledge financial support by the Deutsche Forschungsgemeinschaft (DFG, German Research Foundation) through the Würzburg-Dresden Cluster of Excellence ctd.qmat – Complexity, Topology and Dynamics in Quantum Matter (EXC 2147, project-id 390858490).
AP gratefully acknowledges the Heisenberg Program of the
DFG  (project-id 435041839).
\end{acknowledgments}

\begin{widetext}
\appendix
\section{Far-field expressions}\label{appendix far-field}
In the main text, we have mostly considered far-field expressions for the intensity and power spectrum.
We start from the electric-field operator
\begin{equation}
    \vec{E}(\vec{r}, \tau) = \mu_0\omega_0^2 \sum_{i=1}^N G(\vec{r}-\vec{r}_i, \omega_0) \cdot \vec{d}_i \sigma_i(\tau),
\end{equation}
in which the Green's function $G(\vec{R}_i, \omega_0)$ with $\vec{R}_i = \vec{r} - \vec{r}_i$ appears.
In the far-field limit $r \gg \lambda_0$ with $r = \abs{\vec{r}}$, only terms up to linear order in $1/r$ are kept, allowing us to approximate
\begin{equation}
    G(\vec{R}_i, \omega_0) \approx \frac{\eu^{\imi k_0 R_i}}{4\pi R_i} (\mathbbm{1} - \hat{R}_i\otimes \hat{R}_i).
\end{equation}
Since $R_i = r\sqrt{1 + (r_i/r)^2 - 2\hat{r}\cdot\vec{r}_i/r} \approx r(1 - \hat{r}\cdot\vec{r}_i /r)$, where we used the Taylor expansion of the square root $\sqrt{1 + x} \approx 1 + x/2$, and similarly $1/R_i \approx 1/r$, we obtain
\begin{equation}
    G(\vec{R}_i, \omega_0) \approx \frac{\eu^{\imi k_0 r} \eu^{-\imi k_0 \hat{r}\cdot \vec{r}_i}}{4\pi r} (\mathbbm{1} - \hat{r}\otimes\hat{r}).
\end{equation}
With that, the far-field expression for the correlation function reads
\begin{equation}
        \braket{\vec{E}^\dagger(\vec{r}, \tau) \cdot \vec{E}(\vec{r}, 0)}_{\rm ss}
        \approx \qty(\frac{\mu_0 \omega_0^2 d}{4\pi r})^2 (1 - |\hat{r}\cdot\hat{d}|^2) \sum_{i,j=1}^N \eu^{\imi k_0\hat{r}\cdot(\vec{r}_i - \vec{r}_j)} \braket{\sigma_i^\dagger(\tau)\sigma_j}_{\rm ss}.
\end{equation}
Using $\hat{d} = \mqty(0 & 0 & 1)^\mathrm{T}$ and $\hat{r} = \mqty(\cos(\varphi) & \sin(\varphi) & 0)$ as in the main text and the surface element $\dd{A} = r^2 \sin(\vartheta) \dd{\vartheta}\dd{\varphi}$ with $\vartheta = \pi/2$, we obtain the far-field power spectrum
    \begin{equation}
        S(\omega, \varphi) \approx 2\Re \int_0^\infty \sum_{i,j=1}^N \eu^{\imi k_0 (x_i - x_j) \cos(\varphi)} \braket{\sigma_i^\dagger(\tau) \sigma_j}_{\rm ss} \eu^{-\imi\omega \tau} \dd{\tau}
    \end{equation}
and the far-field intensity
    \begin{equation}
        I(\varphi) = \frac{1}{2\pi} \int_{-\infty}^\infty S(\omega, \varphi)\dd{\omega} 
        \approx \sum_{i,j=1}^N \eu^{\imi k_0 (x_i-x_j)\cos(\varphi)} \braket{\sigma_i^\dagger\sigma_j}_{\rm ss},
    \end{equation}
where in each case we have dropped the constant prefactor. 
Performing the same steps for the second-order correlation function, we find that in the far field
\begin{equation}
    g^{(2)}(0) = \frac{\braket{\vec{E}^\dagger(\vec{r})\cdot (\vec{E}^\dagger(\vec{r})\cdot \vec{E}(\vec{r})) \vec{E}(\vec{r})}_{\rm ss}}{\braket{\vec{E}^\dagger(\vec{r}) \cdot \vec{E}(\vec{r})}_{\rm ss}^2}
    \approx \frac{\sum_{i,j,k,\ell=1}^N \eu^{\imi k_0 (x_i + x_j - x_k - x_\ell)\cos(\varphi)} \braket{\sigma_i^\dagger\sigma_j^\dagger \sigma_k\sigma_\ell}_{\rm ss}}{\qty(\sum_{n,m=1}^N \eu^{\imi k_0 (x_n - x_m)\cos(\varphi)} \braket{\sigma_n^\dagger\sigma_m}_{\rm ss})^2}.
\end{equation}

\section{Edge-state population}\label{appendix edge states}
\begin{figure}
    \centering
    \includegraphics[width=\linewidth]{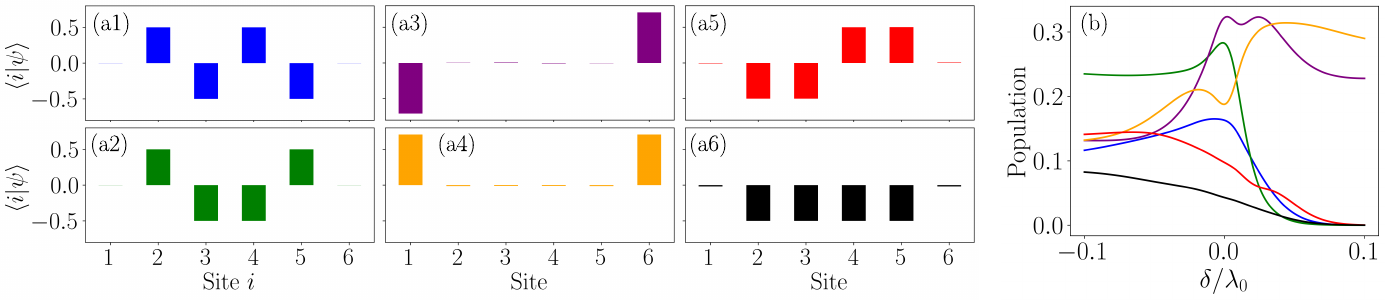}
    \caption{(a1)-(a6) Eigenstates $\ket{\psi}$ of the dipole-dipole-coupling matrix $\Omega_{ij}$ for $N=6$ emitters with $a=0.15\lambda_0$ and $\delta=0.1\lambda_0$. 
    (Gauge is chosen such that $\braket{i|\psi}$ is purely real.)
    The states in (a3) and (a4) are the two edge modes.
    (b) Eigenstate populations for varying $\delta$. 
    The colors correspond to the colors in (a1)-(a6).
    In all panels $R_0=\Gamma_0$.
    }
    \label{fig figure 6}
\end{figure}
In the main text, we have argued that our pumping scheme, in which only the first emitter is incoherently pumped, drives the system into a steady state which is dominated by the two topological edge modes. 
In Figs.~\ref{fig figure 6}(a1)-(a6), we show the eigenstates of the dipole-dipole-coupling-matrix $\Omega_{ij}$ for $N=6$ emitters in the topological phase. 
The two states in panels (a3) and (a4) are the topological edge modes, which are approximately  given by 
\begin{equation}
    \ket{\psi_\pm} \approx \frac{1}{\sqrt{2}} (\ket{N} \pm \ket{1}).
\end{equation}
Therefore, the excited state of the first emitter can be written as a superposition of the two edge modes, $\ket{1} \approx \frac{1}{2} ( \ket{\psi_+} - \ket{\psi_-})$.
This explains why our system shows near-perfect single-photon emission (since only the first emitter is excited) while at the same time revealing interference effects in the radiation pattern (due to the simultaneous participation of two edge states).

The edge state population can be seen explicitly by considering the matrix $u_{ij} = \braket{\sigma_i^\dagger\sigma_j}_{\rm ss}$ whose diagonal elements are the populations of the sites.
Let $T$ be the basis transformation matrix from the site basis into the eigenbasis in which the dipole-dipole-coupling matrix $\Omega_{ij}$ is diagonal. 
Then the diagonal elements of the transformed matrix $\Tilde{u} = T^{-1} u T$ can be interpreted as the populations of the eigenstates.
We show these populations in Fig.~\ref{fig figure 6}(b) for varying $\delta$.
One can see that while in the trivial phase $\delta<0$ all eigenstates contribute to the steady state, in the topological phase, except for an intermediate phase $0<\delta<0.05\lambda_0$, the steady state is dominated by the two topological edge states.

\end{widetext}

\bibliography{bib.bib}% Produces the bibliography via BibTeX.

@PREAMBLE{
 "\providecommand{\noopsort}[1]{}" 
 # "\providecommand{\singleletter}[1]{#1}%" 
}

@book{Schlosshauer_book,
title = {{Decoherence and the Quantum-To-Classical Transition}},
author = {Maximilian Schlosshauer},
publisher = {Springer Berlin, Heidelberg},
year = {2007},
doi = {https://doi.org/10.1007/978-3-540-35775-9},
}

@book{Breuer_book,
title = {{The Theory of Open Quantum Systems}},
author = {Heinz-Peter Breuer and Francesco Petruccione},
publisher = {Oxford},
year = {2007},
doi = {https://doi.org/10.1093/acprof:oso/9780199213900.001.0001},
}

@book{Hecht_book,
title = {{Principles of Nano-Optics}},
author = {Lukas Novotny and Bert Hecht},
publisher = {Cambride University Press},
year = {2012},
doi = {https://doi.org/10.1017/CBO9780511794193},
}

@book{Glauber_book,
title = {{Quantum Theory of Optical Coherence}},
author = {Roy J. Glauber},
publisher = {Wiley-VCH},
year = {2006},
doi = {10.1002/9783527610075},
}

@book{Carmichael_book,
title = {{An Open Systems Approach to Quantum Optics}},
author = {Howard Carmichael},
publisher = {Springer Berlin, Heidelberg},
year = {2009},
doi = {https://doi.org/10.1007/978-3-540-47620-7},
}

@article{Lounis2005,
    author = {B. Lounis and M. Orrit},
    title = {Single-photon sources},
    journal = {Rep. Prog. Phys.},
    volume = {68},
    pages = {1129},
    year = {2005},
    doi = {10.1088/0034-4885/68/5/R04}
}

@article{Obrien2007,
author = {Jeremy L. O'Brien },
title = {Optical Quantum Computing},
journal = {Science},
volume = {318},
pages = {1567},
year = {2007},
doi = {10.1126/science.1142892}
}

@article{Scarani2009,
  title = {The security of practical quantum key distribution},
  author = {Scarani, Valerio and Bechmann-Pasquinucci, Helle and Cerf, Nicolas J. and Du\ifmmode \check{s}\else \v{s}\fi{}ek, Miloslav and L\"utkenhaus, Norbert and Peev, Momtchil},
  journal = {Rev. Mod. Phys.},
  volume = {81},
  issue = {3},
  pages = {1301--1350},
  numpages = {0},
  year = {2009},
  month = {Sep},
  publisher = {American Physical Society},
  doi = {10.1103/RevModPhys.81.1301},
  url = {https://link.aps.org/doi/10.1103/RevModPhys.81.1301}
}

@article{Pirandola2020,
    author = {S. Pirandola and U. L. Andersen and L. Banchi and M. Berta and D. Bunandar and R. Colbeck and D. Englund and T. Gehring and C. Lupo and C. Ottaviani and J. L. Pereira and M. Razavi and J. Shamsul Shaari and M. Tomamichel and V. C. Usenko and G. Vallone and P. Villoresi and P. Wallden},
    title = {Advances in quantum cryptography},
    journal = {Adv. Opt. Photon.},
    volume = {12},
    pages = {1012},
    year = {2020},
    doi = {https://doi.org/10.1364/AOP.361502}
}

@article{Azuma2023,
  title = {Quantum repeaters: From quantum networks to the quantum internet},
  author = {Azuma, Koji and Economou, Sophia E. and Elkouss, David and Hilaire, Paul and Jiang, Liang and Lo, Hoi-Kwong and Tzitrin, Ilan},
  journal = {Rev. Mod. Phys.},
  volume = {95},
  issue = {4},
  pages = {045006},
  numpages = {66},
  year = {2023},
  month = {Dec},
  publisher = {American Physical Society},
  doi = {10.1103/RevModPhys.95.045006},
  url = {https://link.aps.org/doi/10.1103/RevModPhys.95.045006}
}

@article{Ding2016,
  title = {{On-Demand Single Photons with High Extraction Efficiency and Near-Unity Indistinguishability from a Resonantly Driven Quantum Dot in a Micropillar}},
  author = {Ding, Xing and He, Yu and Duan, Z.-C. and Gregersen, Niels and Chen, M.-C. and Unsleber, S. and Maier, S. and Schneider, Christian and Kamp, Martin and H{\"o}fling, Sven and Lu, Chao-Yang and Pan, Jian-Wei},
  journal = {Phys. Rev. Lett.},
  volume = {116},
  pages = {020401},
  year = {2016},
  doi = {10.1103/PhysRevLett.116.020401}
}

@article{Hornung2024,
    author = {Hornung, Florian and Pfister, Ulrich and Bauer, Stephanie and Cyrlyson’s, Dee Rocking and Wang, Dongze and Vijayan, Ponraj and Garcia, Ailton J., Jr. and Covre da Silva, Saimon Filipe and Jetter, Michael and Portalupi, Simone L. and Rastelli, Armando and Michler, Peter},
    title = {{Highly Indistinguishable Single Photons from Droplet-Etched GaAs Quantum Dots Integrated in Single-Mode Waveguides and Beamsplitters}},
    journal = {Nano Letters},
    volume = {24},
    pages = {1184},
    year = {2024},
    doi = {10.1021/acs.nanolett.3c04010},
}

@article{Kagan2021,
    author = {Cherie R. Kagan and Lee C. Bassett and Christopher B. Murray and Sarah M. Thompson},
    title = {{Colloidal Quantum Dots as Platforms for Quantum Information Science}},
    journal = {Chem. Rev.},
    volume = {121},
    pages = {3186},
    year = {2021},
    doi = {https://doi.org/10.1021/acs.chemrev.0c00831}
}

@article{Lodahl2015,
  title = {Interfacing single photons and single quantum dots with photonic nanostructures},
  author = {Lodahl, Peter and Mahmoodian, Sahand and Stobbe, S\o{}ren},
  journal = {Rev. Mod. Phys.},
  volume = {87},
  pages = {347},
  year = {2015},
  doi = {10.1103/RevModPhys.87.347},
}

@article{Kimble1977,
  title = {{Photon Antibunching in Resonance Fluorescence}},
  author = {Kimble, H. J. and Dagenais, M. and Mandel, L.},
  journal = {Phys. Rev. Lett.},
  volume = {39},
  pages = {691},
  year = {1977},
  doi = {10.1103/PhysRevLett.39.691}
}

@article{Kurtsiefer2000,
  title = {{Stable Solid-State Source of Single Photons}},
  author = {Kurtsiefer, Christian and Mayer, Sonja and Zarda, Patrick and Weinfurter, Harald},
  journal = {Phys. Rev. Lett.},
  volume = {85},
  pages = {290},
  year = {2000},
  doi = {10.1103/PhysRevLett.85.290}
}

@article{Aharonovich2011,
    author = {I. Aharonovich and S. Castelletto and D. A. Simpson and C.-H. Su and A. D. Greentree and S. Prawer},
    title = {Diamond-based single-photon emitters},
    journal = {Rep. Prog. Phys.},
    volume = {74},
    pages = {076501},
    year = {2011},
    doi = {10.1088/0034-4885/74/7/076501}
}

@article{Brunel1999,
  title = {{Triggered Source of Single Photons based on Controlled Single Molecule Fluorescence}},
  author = {Brunel, Christian and Lounis, Brahim and Tamarat, Philippe and Orrit, Michel},
  journal = {Phys. Rev. Lett.},
  volume = {83},
  pages = {2722},
  year = {1999},
  publisher = {American Physical Society},
  doi = {10.1103/PhysRevLett.83.2722}
}

@article{Vajner2022,
    author = {D.~A.~Vajner and L.~Rickert and T.~Gao and K.~Kaymazlar and T.~Heindel},
    title = {{Quantum Communication Using Semiconductor Quantum Dots}},
    journal = {Adv. Quantum Technol.},
    volume = {5},
    pages = {2100116},
    year = {2022},
    doi = {https://doi.org/10.1002/qute.202100116}
}

@article{Keni2025,
author = {Arya Keni and Kinjol Barua and Khabat Heshami and Alisa Javadi and Hadiseh Alaeian},
journal = {Opt. Mater. Express},
number = {4},
pages = {626},
publisher = {Optica Publishing Group},
title = {{Single-photon generation: materials, techniques, and the Rydberg exciton frontier [Invited]}},
volume = {15},
year = {2025},
doi = {10.1364/OME.549582}
}

@article{Campaioli2024,
  title = {Quantum Master Equations: Tips and Tricks for Quantum Optics, Quantum Computing, and Beyond},
  author = {Campaioli, Francesco and Cole, Jared H. and Hapuarachchi, Harini},
  journal = {PRX Quantum},
  volume = {5},
  issue = {2},
  pages = {020202},
  numpages = {33},
  year = {2024},
  month = {Jun},
  publisher = {American Physical Society},
  doi = {10.1103/PRXQuantum.5.020202},
  url = {https://link.aps.org/doi/10.1103/PRXQuantum.5.020202}
}

@article{Lehmberg1970I,
  title = {{Radiation from an $N$-Atom System. I. General Formalism}},
  author = {Lehmberg, R. H.},
  journal = {Phys. Rev. A},
  volume = {2},
  pages = {883},
  year = {1970},
  doi = {10.1103/PhysRevA.2.883}
}

@article{Lehmberg1970II,
  title = {{Radiation from an $N$-Atom System. II. Spontaneous Emission from a Pair of Atoms}},
  author = {Lehmberg, R. H.},
  journal = {Phys. Rev. A},
  volume = {2},
  pages = {889},
  year = {1970},
  doi = {10.1103/PhysRevA.2.889}
}

@article{Jen2025,
    author = {H. H. Jen},
    title = {Photon-mediated dipole–dipole interactions as a resource for quantum science and technology in cold atoms},
    journal = {Quantum Sci. Technol.},
    volume = {10},
    pages = {023001},
    year = {2025},
    doi = {10.1088/2058-9565/adbb86}
}

@article{Wang2018,
  title = {{Topological photonic states in one-dimensional dimerized ultracold atomic chains}},
  author = {Wang, B. X. and Zhao, C. Y.},
  journal = {Phys. Rev. A},
  volume = {98},
  issue = {2},
  pages = {023808},
  numpages = {12},
  year = {2018},
  month = {Aug},
  publisher = {American Physical Society},
  doi = {10.1103/PhysRevA.98.023808},
  url = {https://link.aps.org/doi/10.1103/PhysRevA.98.023808}
}

@article{Sturm2025,
  title = {Polarization-dependent topology in quantum emitter chains},
  author = {Sturm, Jonathan and P{\'a}lffy, Adriana},
  journal = {Phys. Rev. Res.},
  volume = {7},
  pages = {L032069},
  year = {2025},
  doi = {10.1103/3t6s-fnsq}
}

@article{Bauer2018,
  title = {{High-Harmonic Generation in Solids with and without Topological Edge States}},
  author = {Bauer, Dieter and Hansen, Kenneth K.},
  journal = {Phys. Rev. Lett.},
  volume = {120},
  pages = {177401},
  year = {2018},
  doi = {10.1103/PhysRevLett.120.177401}
}

@article{Dicke1954,
    author = {R. H. Dicke},
    title = {{Coherence in Spontaneous Radiation Processes}},
    journal = {Phys. Rev.},
    volume = {93},
    pages = {99},
    year = {1954},
    doi = {https://doi.org/10.1103/PhysRev.93.99}
}

@article{Bychek2025,
  title = {{Nanoscale Mirrorless Superradiant Lasing}},
  author = {Bychek, Anna and Holzinger, Raphael and Ritsch, Helmut},
  journal = {Phys. Rev. Lett.},
  volume = {135},
  pages = {143601},
  year = {2025},
  doi = {10.1103/rbs2-2pd5}
}

@article{Tonoyan2026,
  title = {Second-order correlations in directed emissions in sodium atoms},
  author = {Tonoyan, Ara and Sahoo, Sushree Subhadarshinee and Gogyan, Anahit and Tretiak, Oleg and Aramyan, Razmik and Akulshin, Alexander and Budker, Dmitry},
  journal = {Phys. Rev. A},
  volume = {113},
  pages = {063710},
  year = {2026},
  doi = {10.1103/q1s4-njnb}
}

@article{Holzinger2020,
  title = {Nanoscale Coherent Light Source},
  author = {Holzinger, Raphael and Plankensteiner, David and Ostermann, Laurin and Ritsch, Helmut},
  journal = {Phys. Rev. Lett.},
  volume = {124},
  pages = {253603},
  year = {2020},
  doi = {10.1103/PhysRevLett.124.253603}
}

@article{Garcia2017,
  title = {{Exponential Improvement in Photon Storage Fidelities Using Subradiance and ``Selective Radiance'' in Atomic Arrays}},
  author = {Asenjo-Garcia, A. and Moreno-Cardoner, M. and Albrecht, A. and Kimble, H. J. and Chang, D. E.},
  journal = {Phys. Rev. X},
  volume = {7},
  pages = {031024},
  year = {2017},
  doi = {10.1103/PhysRevX.7.031024}
}

@article{Dung2002,
  title = {{Resonant dipole-dipole interaction in the presence of dispersing and absorbing surroundings}},
  author = {Dung, Ho Trung and Kn\"oll, Ludwig and Welsch, Dirk-Gunnar},
  journal = {Phys. Rev. A},
  volume = {66},
  issue = {6},
  pages = {063810},
  numpages = {16},
  year = {2002},
  month = {Dec},
  publisher = {American Physical Society},
  doi = {10.1103/PhysRevA.66.063810},
  url = {https://link.aps.org/doi/10.1103/PhysRevA.66.063810}
}

@article{Reitz2022,
  title = {{Cooperative Quantum Phenomena in Light-Matter Platforms}},
  author = {Reitz, Michael and Sommer, Christian and Genes, Claudiu},
  journal = {PRX Quantum},
  volume = {3},
  issue = {1},
  pages = {010201},
  numpages = {33},
  year = {2022},
  month = {Jan},
  publisher = {American Physical Society},
  doi = {10.1103/PRXQuantum.3.010201},
  url = {https://link.aps.org/doi/10.1103/PRXQuantum.3.010201}
}

@article{Wang2024,
    author = {Yubin Wang and Huawen Xu and Xinyi Deng and Timothy C. H. Liew and Sanjib Ghosh and Qihua Xiong},
    title = {{Topological single-photon emission from quantum emitter chains}},
    journal = {npj Quantum Inf},
    volume = {10},
    pages = {13},
    year = {2024},
    doi = {https://doi.org/10.1038/s41534-024-00807-y}
}

@misc{quantumpotics,
    note = {Numerical simulations have been performed using the QuantumOptics.jl package: S.~Krämer, D.~Plankensteiner, L.~Ostermann, and H.~Ritsch, QuantumOptics.jl: A Julia framework for simulating open quantum systems, \href{https://doi.org/10.1016/j.cpc.2018.02.004}{Computer Physics Communications \textbf{227}, 109 (2018)}}
}

@article{kong2025,
title = {Recent advances of photolithography patterning of quantum dots for micro-display applications},
journal = {Nano Materials Science},
volume = {7},
pages = {49},
year = {2025},
doi = {https://doi.org/10.1016/j.nanoms.2024.03.005},
author = {Xuemin Kong and Xiaotong Fan and Yuhui Wang and Yunshu Luo and Yihang Chen and Tingzhu Wu and Zhong Chen and Yue Lin and Shuli Wang}
}

@article{Xie2015,
    author = {Xie, Weiqiang and Gomes, Raquel and Aubert, Tangi and Bisschop, Suzanne and Zhu, Yunpeng and Hens, Zeger and Brainis, Edouard and Van Thourhout, Dries},
    title = {{Nanoscale and Single-Dot Patterning of Colloidal Quantum Dots}},
    journal = {Nano Letters},
    volume = {15},
    pages = {7481},
    year = {2015},
    doi = {10.1021/acs.nanolett.5b03068}
}

@article{Curto2010,
author = {Alberto G. Curto  and Giorgio Volpe  and Tim H. Taminiau  and Mark P. Kreuzer  and Romain Quidant  and Niek F. van Hulst },
title = {{Unidirectional Emission of a Quantum Dot Coupled to a Nanoantenna}},
journal = {Science},
volume = {329},
pages = {930},
year = {2010},
doi = {10.1126/science.1191922}
}

@article{Seo2018,
  author  = {Seo, In Cheol and Woo, Byung Hoon and An, Soo-Chan and
             Lee, Eunsongyi and Jeong, Hoon Yeub and Lim, Yeonsoo and
             Jun, Young Chul},
  title   = {{Electron-Beam-Induced Nanopatterning of J-Aggregate Thin Films for Excitonic and Photonic Response Control}},
  journal = {Advanced Optical Materials},
  year    = {2018},
  volume  = {6},
  pages   = {1800583},
  doi     = {10.1002/adom.201800583}
}

@article{Schurr2025,
  author  = {Schurr, Benedikt and Hensen, Matthias and
             Brenneis, Luisa and Kessler, Philipp and Qin, Jin and
             Lisinetskii, Victor and Thomale, Ronny and
             Brixner, Tobias and Hecht, Bert},
  title   = {{Plasmonic {Su--Schrieffer--Heeger} chains with strong coupling amplitudes}},
  journal = {Science Advances},
  year    = {2025},
  volume  = {11},
  pages   = {eaea3844},
  doi     = {10.1126/sciadv.aea3844}
}

@article{Su1979,
  title = {{Solitons in Polyacetylene}},
  author = {Su, W. P. and Schrieffer, J. R. and Heeger, A. J.},
  journal = {Phys. Rev. Lett.},
  volume = {42},
  issue = {25},
  pages = {1698--1701},
  numpages = {0},
  year = {1979},
  month = {Jun},
  publisher = {American Physical Society},
  doi = {10.1103/PhysRevLett.42.1698},
  url = {https://link.aps.org/doi/10.1103/PhysRevLett.42.1698}
}

@book{Asboth_book,
title = {{A Short Course on Topological Insulators}},
author = {János K. Asbóth and László Oroszlány and András Pályi},
publisher = {Springer Cham},
year = {2016},
doi = {https://doi.org/10.1007/978-3-319-25607-8},
}

@article{Wuerthner2011,
author = {Würthner, Frank and Kaiser, Theo E. and Saha-Möller, Chantu R.},
title = {J-Aggregates: From Serendipitous Discovery to Supramolecular Engineering of Functional Dye Materials},
journal = {Angew. Chem. Int. Ed.},
volume = {50},
number = {15},
pages = {3376-3410},
doi = {https://doi.org/10.1002/anie.201002307},
url = {https://onlinelibrary.wiley.com/doi/abs/10.1002/anie.201002307},
year = {2011}
}

@article{Hestand2018,
    author = {Hestand, Nicholas J. and Spano, Frank C.},
    title = {Expanded Theory of {H}- and {J}‑Molecular Aggregates: {T}he Effects of Vibronic Coupling and Intermolecular Charge Transfer},
    journal = {Chem. Rev.},
    volume = {118},
    number = {15},
    pages = {7069-7163},
    year = {2018},
    month = {08},
    issn = {0009-2665},
    doi = {10.1021/acs.chemrev.7b00581},
    url = {https://doi.org/10.1021/acs.chemrev.7b00581}
}

\end{document}